\documentclass[
aps,
prl,
twocolumn,
nofootinbib,
longbibliography
]{revtex4-2}

\usepackage{hyperref}
\usepackage{url}
\usepackage{setspace}
\usepackage{amsmath,amssymb,amscd,braket}
\usepackage{amsfonts}
\usepackage{color}
\usepackage{graphicx}
\usepackage{xcolor}

\newcommand{\bea}{\begin{eqnarray}}
\newcommand{\eea}{\end{eqnarray}}
\newcommand{\be}{\begin{equation}}
\newcommand{\ee}{\end{equation}}
\newcommand{\ba}{\begin{align}}
\newcommand{\ea}{\end{align}}

\begin{document}
\title{Bouncing singularities in Schwarzschild:\\
a geometric origin of the QNM convergence region 
}

\author{Paolo Arnaudo}
\email{p.arnaudo@soton.ac.uk}
\affiliation{Mathematical Sciences and STAG Research Centre, University of Southampton, Highfield, Southampton SO17 1BJ, UK}

\author{Benjamin Withers}
\email{b.s.withers@soton.ac.uk}
\affiliation{Mathematical Sciences and STAG Research Centre, University of Southampton, Highfield, Southampton SO17 1BJ, UK}

\begin{abstract}
We show analytically that the convergence of the QNM expansion of the retarded Green's function of the Schwarzschild spacetime is set by a singularity in the complex time plane. The singularity has a simple geometric origin: it is an example of a `bouncing singularity' in the language of AdS/CFT literature, caused by a null geodesic which bounces from the black hole singularity. Our work explains why the QNM convergence region at real times is bounded by null ray which scatters from the gravitational potential at a seemingly unremarkable point ($r_* = 0$ in the conventions of previous work) -- this ray is the same distance from the origin as the bouncing singularity in the relevant complex plane. The same set of singularities are responsible for an annular region of convergence for the Matsubara mode sum which describes the early time behaviour of the Schwarzschild Green's function for perturbations close to the horizon.
\end{abstract}

\maketitle

\section{Introduction}
With recent advances in gravitational wave astronomy it is increasingly important to gain precise theoretical control on the ringdown behaviour of black holes. Indeed, through observations of the loudest event to date, GW250114, it has been possible to constrain the first few quasinormal modes (QNMs) of the remnant black hole \cite{LVKloud} highlighting the increasing relevance of the growing field of black hole spectroscopy \cite{Berti:2025hly}.

In this work we revisit the retarded Green's function, $G_R$, for linear perturbations of the Schwarzschild black hole. Working at fixed spin $s$ and angular quantum numbers $m, \ell$, at late enough times the Green's function can be expressed as a branch cut contribution, plus a convergent sum over QNMs. Writing the QNM part as
\be
G^\text{QNM}_R(t, t', r, r') = \sum_{n=0}^\infty c_n(r,r') e^{-i\omega_n (t-t')}, \label{QNMsum}
\ee
the question this letter answers is simple: what physical feature sets the region of convergence of the QNM sum \eqref{QNMsum}?

\begin{figure}[h!]
\centering
\includegraphics[width=\columnwidth]{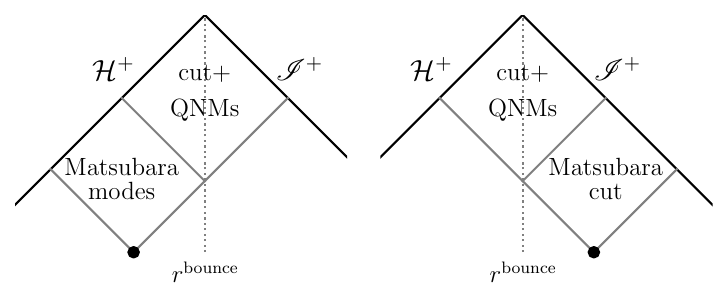}
\caption{The spectral decomposition of \cite{Arnaudo:2025uos} in the case of the Schwarzschild spacetime. The black dot corresponds to the delta function source for the retarded Green's function, whose spectral decomposition depends on the region of spacetime, as illustrated. The regions are determined by a radius $r=r^\text{bounce}$ where the lightcone scatters from the black hole potential, given by \eqref{rbounce}. The goal of this work is to explain the physical origin of this location.}
\label{fig:decomposition}
\end{figure}

A complete decomposition of $G_R$ was provided in our earlier work \cite{Arnaudo:2025uos}, extending the seminal spectral decomposition of \cite{Leaver} which diverges at early times. To resolve this, our construction showed how the Green's function must be split into multiple parts together with a contour prescription. For related earlier work see \cite{Andersson:1996cm}. The choice of contour is conditional on a partitioning of spacetime according to the scattering of the Green's function lightcone from the black hole potential at a specific radius, $r^\text{bounce}$.\footnote{At the same time, an analogous construction was given in \cite{Kuntz:2025gdq} for perturbations of the P\"oschl-Teller potential.} The decomposition is illustrated in FIG. \ref{fig:decomposition}. Therefore given $t', r'$, then $r^\text{bounce}$ determines the convergence region for \eqref{QNMsum}.

The value for $r^\text{bounce}$ was first approximated in \cite{Arnaudo:2025uos} using an instanton expansion of Heun connection coefficients. In our accompanying work \cite{Arnaudo:2025kit} it was computed directly through convergence analysis of the branch cut contribution, finding,
\be
r^\text{bounce} = 2M\left(1 + W_0\left(1/e\right)\right), \label{rbounce}
\ee
where $W_k(z)$ is the Lambert W function. This is consistent with the convergence seen in subsequent approximate numerical treatments by \cite{Su:2026fvj, Ma:2026qbq}.\footnote{See also \cite{DeAmicis:2025xuh} which explores the QNM convergence region, but which finds differing results.}

More compactly, using the tortoise coordinate
\be
r_* = r + 2M \log\left(\frac{r}{2M}-1\right) + C, \label{tortoise}
\ee
where $C$ is an arbitrary integration constant, 
the bounce radius \eqref{rbounce} can be expressed as
\be
r^\text{bounce}_* = C. \label{rstarbounce}
\ee
As a matter of convention, in previous work the integration constant was taken to be $C=0$ thus establishing $r^\text{bounce}_* = 0$ as the relevant shorthand. For maximum clarity, in this work we keep $C$ arbitrary. Since physical quantities do not depend on the arbitrary integration constant $C$, we will find that where $r_*$ appears in them it will do so in the combination $r_*-C$.

The key issue that our work addresses is that \eqref{rbounce}/\eqref{rstarbounce} does not appear to correspond to a location of any physical significance. For instance, it is not the peak of the potential or the light ring. We will show that this location carries a privileged geometrical status, despite its curious numerical value. A hint that this is true can already be seen by turning to Kruskal–Szekeres coordinates,
\begin{equation}
U=-e^{-\frac{t-r_*}{4M}},\qquad V=e^{\frac{t+r_*}{4M}}.
\label{eq:kruskal_I}
\end{equation}
In these coordinates, the location of the black hole singularity $r=0$ is given by the locus
\begin{equation}\label{BHsingKSC}
UV=e^{\frac{C}{2M}},
\end{equation}
while the bounce location \eqref{rbounce} corresponds simply to a reflection of \eqref{BHsingKSC} in the Kruskal plane,
\begin{equation}\label{bounceKSC}
UV =-e^{\frac{C}{2M}}.
\end{equation}
This hints at a deeper geometrical origin for the reason behind \eqref{rbounce}.

Indeed, we show that the true origin of \eqref{rbounce} is geometrical in the form of a pair of so-called `bouncing singularities' in the complex $t$-plane,
\bea
-t+r_*+t'+r_*' &=& 2C+i\beta\left(n-\frac{1}{2}\right),\quad \text{(future)}\quad\label{futurebounce}\\
t+r_*-t'+r_*' &=& 2C+i\beta\left(n-\frac{1}{2}\right), \quad \text{(past)}\quad\label{pastbounce}
\eea
for $n\in\mathbb Z$, $C$ is the integration constant in \eqref{tortoise} and where $\beta = 8\pi M$, the inverse Hawking temperature of the black hole. These singularities appear because they are loci in spacetime which are connected to $t',r'$ by null geodesics which bounce from the future/past black hole singularity. Bouncing singularities is a development drawn from the AdS/CFT literature, where such singularities appear in analytic continuations of thermal correlation functions \cite{Fidkowski:2003nf, Festuccia:2005pi} and for which there has been considerable recent interest \cite{Parisini:2023nbd, Ceplak:2024bja, Buric:2025anb, Barrat:2025nvu, Buric:2025fye, Barrat:2025twb, Afkhami-Jeddi:2025wra, Dodelson:2025jff, Ceplak:2025dds, Jia:2025jbi, Arnaudo:2026der, Grozdanov:2026cut, Jia:2026ryl}.

We show that the bouncing singularities \eqref{futurebounce}, \eqref{pastbounce} set the radius of convergence for the QNM sum, \eqref{QNMsum}. Which singularity is relevant depends on the location of $r_*'$. If $r_*' > C$ then \eqref{futurebounce} sets the radius of convergence, while if $r_*' < C$ then \eqref{pastbounce} sets the radius of convergence. These singularities do not appear in the physical spacetime, but, they do determine the region of convergence of the QNM sum in the physical spacetime. The boundary of convergence is precisely the lightcone of the retarded Green's function scattering from the potential at \eqref{rbounce}, thus establishing the main result.

In addition to the QNM sum, the same set of singularities are responsible for the convergence region of the Matsubara mode sum near the horizon, as in the left panel of FIG. \ref{fig:decomposition}. In this case the Matsubara mode sum is a Laurent series with an annular region of convergence, corresponding to the region between two singularities. One singularity is \eqref{pastbounce} and the other is the ingoing component of the lightcone singularity. By performing the asymptotic sum of Matsubara modes explicitly, we recover functions which have precisely these singularities.
\\\\
The rest of the letter is structured as follows. We first give the basic definitions and conventions for linear spin-$s$ perturbations of Schwarzschild and the retarded Green's function. We then explain the existence of bouncing geodesics in Schwarzschild spacetime, before showing how they correspond to the complex time singularities \eqref{futurebounce} and \eqref{pastbounce}, and how they determine the QNM radius of convergence. Finally we show that the early-time near-horizon decomposition of the Green's function in Matsubara modes also has an annular region of convergence bounded by these singularities.

\section{Schwarzschild perturbations and the radial Green function}

We consider a class of massless linear perturbations with spin $s$ around the four-dimensional Schwarzschild black hole,
\begin{equation}
ds^2=-f(r)\,\mathrm{d}t^2+\frac{\mathrm{d}r^2}{f(r)}+r^2\,\mathrm{d}\Omega_2^2,
\end{equation}
with $f(r)=1-\frac{2M}{r}$.
The spin label is $s=0,1,2$, corresponds respectively to scalar, electromagnetic, and vector-type (Regge-Wheeler) gravitational perturbations, which are decomposed into Fourier modes,
\begin{equation}
\Phi(t,r,\theta,\varphi)=\int \mathrm{d}\omega\sum_{\ell,m}e^{-i\omega t}Y_{\ell m}(\theta,\varphi)\frac{\phi(r)}{r}.
\end{equation}
The resulting radial equation is
\begin{equation}\label{reggewheeler}
\phi''(r)+\frac{f'(r)}{f(r)}\phi'(r)+\frac{\omega^2-V_s(r)}{f(r)^2}\phi(r)=0,
\end{equation}
with the potential
\begin{equation}\label{rw_potential}
V_s(r)=f(r)\left[\frac{\ell(\ell+1)}{r^2}+(1-s^2)\frac{2M}{r^3}\right].
\end{equation}
We denote by $\phi_{\text{in}}$ the solution which is ingoing at the future horizon, and by $\phi_{\text{up}}$ the solution which is outgoing at future null infinity:
\begin{equation}
\begin{aligned}
\phi_{\text{in}}(r)&\sim e^{-i\omega r_*},\qquad r\to 2M,\\
\phi_{\text{up}}(r)&\sim e^{i\omega r_*},\qquad r\to\infty.
\end{aligned}
\end{equation}
From these homogeneous solutions, the frequency-domain retarded Green's function is constructed
\be
\widetilde{G}_R(\omega, r, r') = \frac{1}{\mathcal{W}} \times
\begin{cases}
\phi_\text{in}(r) \, \phi_\text{up}(r'), & r < r' \\
\phi_\text{in}(r') \, \phi_\text{up}(r), & r > r',
\end{cases} \label{Gformal}
\ee
where the Wronskian $\mathcal{W}$ is $\mathcal{W} = \phi_\text{up} \partial_{r_\ast}\phi_\text{in} - \phi_\text{in} \partial_{r_\ast}\phi_\text{up}$. Finally, the time-domain retarded Green's function of interest is obtained by computing the Fourier transform, 
\be
G_R(t,t', r, r') = \int_{-\infty}^\infty \frac{d\omega}{2\pi} \widetilde{G}_R(\omega, r, r') e^{-i \omega (t-t')}, \label{GRFT}
\ee
whose evaluation is the subject of our earlier spectral decomposition paper, \cite{Arnaudo:2025uos}.

\section{Bouncing geodesics}
\begin{figure}[h!]
\centering
\includegraphics[width=0.8\columnwidth]{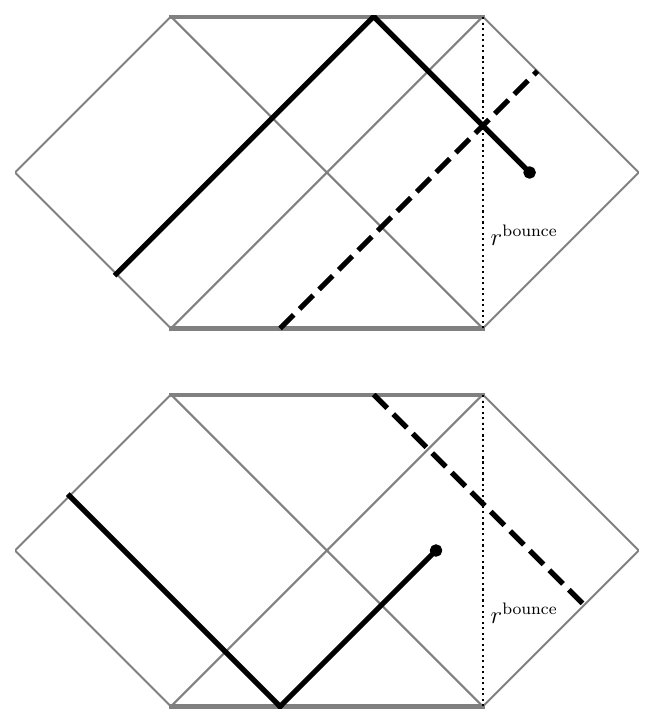}
\caption{Bouncing geodesics in the Schwarzschild spacetime. Consider a spacelike geodesic starting at a point in the right external region (the black dot), and directed radially inwards. There is a large energy limit of the geodesic in which it becomes null and bounces from the singularity (black solid lines), either to the future (\textbf{upper panel}) or the past (\textbf{lower panel}). The black dashed lines correspond to a null geodesic that has the same value of $|X| = \left|e^{-\frac{2\pi}{\beta}t}\right|$ as the geodesics after they bounce from the singularity.}
\label{fig:bounces}
\end{figure}

Of central importance in this work are the existence of so-called `bouncing geodesics' in the Schwarzschild spacetime. These are a limit of spacelike geodesics that connect the two external regions of the maximally extended Schwarzschild spacetime. In this limit, the geodesic becomes null, where it bounces from the black hole singularity, either to the future, or to the past. This is illustrated in FIG. \ref{fig:bounces}.

A radial spacelike geodesic $x^\mu(\lambda)=(t(\lambda),r(\lambda))$, affinely parametrized by $\lambda$, obeys
\begin{equation}\label{normalisationgeo}
-f(r)\dot{t}^2+\frac{\dot{r}^2}{f(r)}=1,
\end{equation}
where the dot denotes derivative with respect to the affine parameter $\lambda$.
Since $\partial_t$ is a Killing vector, the quantity
\begin{equation}\label{tdot}
E=f(r)\dot{t}
\end{equation}
is conserved along the geodesic.
Substituting $\dot{t}=E/f(r)$ into \eqref{normalisationgeo} gives
\begin{equation}
\dot{r}^2=E^2+f(r)=E^2+1-\frac{2M}{r}.
\label{rdot}
\end{equation}
The turning point is determined by $\dot{r} = 0$, so that
\begin{equation}\label{rmin}
r_{\min}(E) = \frac{2M}{1+E^2}.
\end{equation}
It follows immediately that
\begin{equation}\label{rminlimit}
r_{\min}(E) \to 0\qquad\text{as}\qquad E \to \infty.
\end{equation}
Thus the family of radial spacelike geodesics connecting the two exterior regions develops a turning point that approaches the Schwarzschild singularity.

Away from the turning point,
\begin{equation}
\frac{\mathrm{d}t}{\mathrm{d}r}=\frac{\dot{t}}{\dot{r}}=\pm\frac{E}{f(r)\sqrt{E^2+f(r)}}=\pm\frac{1}{f(r)}+O(E^{-2}),
\end{equation}
where we used \eqref{tdot} and \eqref{rdot}.
Thus, each branch becomes a radial null curve in the large-$E$ limit:
\begin{equation}
\frac{\mathrm{d}t}{\mathrm{d}r}=\pm\frac{1}{f(r)}.
\end{equation}
In terms of the tortoise coordinate \eqref{tortoise},  these null curves are
\begin{equation}
u=t-r_*=\mathrm{const},\qquad v=t+r_*=\mathrm{const}.
\end{equation}
The limiting object is therefore a broken radial null curve, made of one branch with $v=\mathrm{const}$ and one branch with $u=\mathrm{const}$, with vertex at the singularity $r=0$.

\begin{figure*}[t!]
\centering
\includegraphics[width=0.8\textwidth]{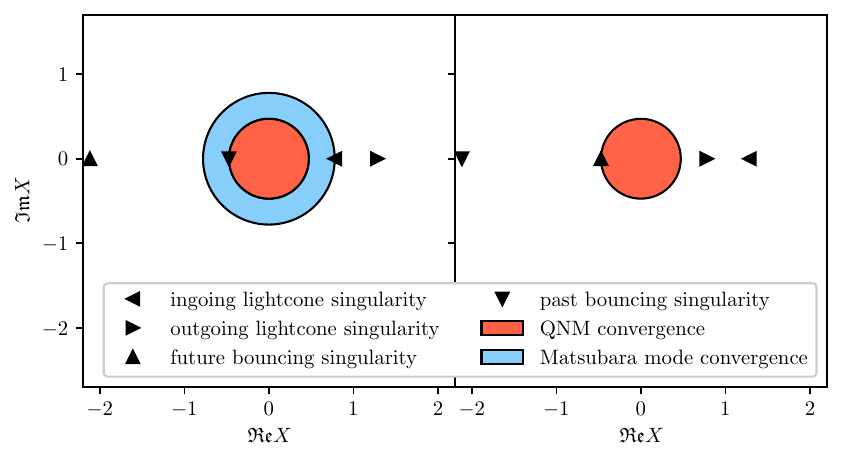}
\caption{Lightcone singularities in the complex $X = e^{-\frac{2\pi}{\beta} t}$ plane for the Schwarzschild spacetime. As shown in the text, these set the spacetime convergence regions of the QNM and Matsubara mode sums for the retarded Green's function. In both cases the QNM convergence region is set by the closest singularity to $X=0$, which occurs for bouncing singularities at complex $t$, i.e. $X<0$. We show two different examples. \textbf{Left panel:} the case where $r_* < r_*' < C$ (plotted using values $r_* = -2$, $r_*'=-1$, $C=0$, $t'=0$ and $M=1$). This corresponds to the left hand panel in FIG. \ref{fig:decomposition}, where the convergence of the QNM sum is set by the past bouncing singularity \eqref{pastbounce} while the Matsubara mode sum is convergent in an annulus. \textbf{Right panel:} the case where $r_* > r_*' > C$ (plotted using values $r_* = 2$, $r_*' = 1$, $C=0$, $t'=0$ and $M=1$). This corresponds to the right hand panel in FIG. \ref{fig:decomposition}, where the convergence of the QNM sum is set by the future bouncing singularity \eqref{futurebounce}.}
\label{fig:Xplanes}
\end{figure*}

\section{QNM convergence}

We now derive the QNM convergence region. Consider the part of the retarded Green's function expressed by a QNM residue sum \eqref{QNMsum}. At large overtone number \cite{Motl:2002hd, Berti:2009kk}
\begin{equation}\label{qnm_asymptotic}
\omega_n\sim -i\frac{2\pi}{\beta} n +\mathcal{O}\left(n^0\right), \qquad n\to \infty.
\end{equation}
Since further subleading corrections in $n$ do not affect the radius of convergence of the residue sum, we can restrict our attention to the series whose coefficients are given by the asymptotic behaviour of those in \eqref{QNMsum}.
Introducing 
\begin{equation}
X=e^{-\frac{2\pi}{\beta}t},\label{Xdef}
\end{equation}
the convergence problem is thus reduced to studying a power series in $X$, whose radius of convergence is fixed by the closest singularity of the (analytically continued) $G_R$ to $X = 0$ in the complex $X$ plane.

Lightcone singularities of $G_R$ are governed by the standard Hadamard propagation of singularities along smooth null geodesics \cite{hadamard, Duistermaat1972, Friedlander1975, Hormander1987} (see also \cite{Casals:2016qyj} for a treatment in the specific context of Schwarzschild black hole, and \cite{Grozdanov:2026cut} for a recent discussion in the context of AdS/CFT). As recent work in AdS/CFT suggests, this includes also bouncing geodesics as broken null limits of spacelike geodesics described in the previous section. These give rise to bouncing singularities at complex $t$ in an analytic continuation of $G_R$, that is, through analytically continuing the Fourier integral \eqref{GRFT} to $t\in \mathbb{C}$.

To derive these points in $X$, let us start by fixing a point in the right exterior region,
\begin{equation}
p_R = (U_R,V_R) \in I,\qquad (U_R<0, V_R>0).
\end{equation}
The ingoing radial null branch from $p_R$ is the line
\begin{equation}
V = V_R. 
\end{equation}
Its intersection with the future singularity \eqref{BHsingKSC} is determined by
\begin{equation}
U_b V_b = e^{\frac{C}{2M}},\qquad V_b = V_R,
\end{equation}
from which one gets the bounced lightray
\begin{equation}
U = \frac{e^{\frac{C}{2M}}}{V_R}.
\end{equation}
Analogously, we can also consider the past directed lightcone bouncing from the past singularity, giving
\begin{equation}
V = \frac{e^{\frac{C}{2M}}}{U_R}.
\end{equation}
Hence, in total we have four singular loci,
\bea
U &=& U_R \;\implies\; X = e^{-\frac{2\pi}\beta\left(t'+r_*-r_*'\right)} \label{sing_in}\\
V &=& V_R \;\implies\; X = e^{-\frac{2\pi}\beta\left(t'-r_*+r_*'\right)} \label{sing_out}\\
U &=& \frac{e^{\frac{C}{2M}}}{V_R} \;\implies\; X = -e^{-\frac{2\pi}\beta\left(t'+r_*+r_*'-2C\right)} \label{sing_future}\\
V &=& \frac{e^{\frac{C}{2M}}}{U_R} \;\implies\; X = -e^{-\frac{2\pi}\beta\left(t'-r_*-r_*'+2C\right)} \label{sing_past}
\eea
corresponding to the ingoing lightcone singularity, outgoing lightcone singularity, future bounced singularity, and past bounced singularity respectively. To compute $X$ the $U,V$ were converted to standard $t, r_*$ coordinates in the right external region using \eqref{eq:kruskal_I}, while $U_R,V_R$ were converted to $t', r_*'$. In the complex $X$ plane these four singularities are shown in FIG. \ref{fig:Xplanes}.

The direct null loci are ordinary Hadamard lightcone singularities associated with smooth radial null geodesics. The bounced loci, instead, should be understood as image singularities associated with the null limit of radial spacelike geodesics whose turning point approaches the black hole or white hole singularity.

The radius of convergence of \eqref{QNMsum} is set by singularity among \eqref{sing_in}, \eqref{sing_out}, \eqref{sing_future}, \eqref{sing_past} closest to $X=0$, which is of bouncing type and sits at $X<0$, as illustrated in FIG. \ref{fig:Xplanes}. The disk of convergence in the complex $X$ plane implies that convergence in the physical spacetime ($X>0$) is given by 
\begin{equation}
t-t'>\max\left\{|r_*-r_*'|,|r_*+r_*'-2C|\right\},
\end{equation}
where the first term represents the direct lightcone, while the second represents the reflection of the lightcone from \eqref{rstarbounce}. This establishes the QNM convergence region depicted in FIG. \ref{fig:decomposition}, where $r^\text{bounce}$ is determined by the bouncing singularity to be the value \eqref{rbounce}.

This result is consistent with the analysis in \cite{Andersson:1996cm, Berti:2006wq}. More precisely, in \cite{Berti:2006wq}, an asymptotic expression for the QNM residues was derived, whose magnitude scale as $1/n$ in the large $n$ regime. This agrees with the numerical computations in \cite{Andersson:1996cm}, where it is also stated that the values of $B_n$ alternate in sign. From \cite{Arnaudo:2025uos} the QNM contributions in this case come from the term proportional to $\phi_\text{up}(r')\phi_\text{up}(r)$ once connection formula are applied, which asymptotically behaves as $e^{i\omega(r_*+r_*'-2C)}$. Together, these statements imply that the sum \eqref{QNMsum} has the asymptotic behaviour
\begin{equation}
\begin{aligned}
\sum_{n>0}\frac{(-1)^n}{n}&\left(X e^{\frac{2\pi}{\beta}(t'+r_*'+r_*-2C)}\right)^n\\
&=-\log (X e^{\frac{2\pi}{\beta}(t'+r_*'+r_*-2C)}+1),
\end{aligned}
\end{equation}
which has a logarithmic singularity at $X=-e^{-\frac{2\pi}{\beta}(t'+r_*'+r_*-2C)}$, matching the location of the future bouncing singularity of this work.

\section{Matsubara mode convergence}
In the previous section we established singularities of the complex $X$ plane, where $X$ is defined in \eqref{Xdef}. These set the convergence region of the QNM sum in the retarded Green's function. Similarly, in this section we will establish that they set the annular region of convergence for the the Matsubara residue sum in the near horizon region. 

To perform the analysis analytically, it is useful to recall the confluent Heun form of the radial equation. Defining $z = r/(2M)$ and
\begin{equation}\label{redefinitionschwarzschild}
\begin{aligned}
\phi(z)=p(z)\,\psi(z),\qquad
p(z)=\sqrt{\frac{z}{z-1}},
\end{aligned}
\end{equation}
the ODE \eqref{reggewheeler} becomes
\begin{equation}\label{RWCHeun}
\resizebox{\columnwidth}{!}{$
\psi''(z)+\frac{1-4\ell(\ell+1)(z-1)z+16M^2\omega^2z^4+4s^2(z-1)}{4(z-1)^2z^2}\psi(z)=0$,}
\end{equation}
which is a confluent Heun equation $\psi''(z)+V_{\text{CH}}(z)\,\psi(z)=0$, with the potential of the form
\begin{equation}\label{canonicalconfluentHeun}
\begin{aligned}
\resizebox{\columnwidth}{!}{$
V_{\text{CH}}(z)=\frac{a_0^2+a_1^2+u-\frac{1}{2}}{z (z-1)}+\frac{\frac{1}{4}-a_0^2}{z^2}+\frac{\frac{1}{4}-a_1^2}{(z-1)^2}-\frac{\epsilon^2}{4}+\frac{\epsilon \mu}{z}$,}
\end{aligned}
\end{equation}
where the dictionary with \eqref{RWCHeun} is given in the supplemental material. In the Heun language, assuming $r<r'$, the retarded Green's function can be written as
\begin{equation}\label{greenretardedz}
\begin{aligned}
\widetilde{G}(\omega, z, z') = &\,2M\frac{p(z)p(z')\psi_{\text{in}}(z)\psi_{\text{up}}(z')}{\psi_{\mathrm{up}}(z) \frac{d}{d z}\psi_{\mathrm{in}}(z) - \psi_{\mathrm{in}}(z) \frac{d}{d z} \psi_{\mathrm{up}}(z)},
\end{aligned}
\end{equation}
where the expressions of $\psi_{\text{in}},\psi_{\text{up}}$ in terms of confluent Heun functions can be found in the supplemental material. Finally, recall that the Matsubara frequencies associated with the horizon temperature are
\begin{equation}\label{MatsubaraSchw}
\omega_k=\frac{ik}{4M},\qquad k\in\mathbb{Z}.
\end{equation}

We start the analysis by considering the connection problem expressing the solution outgoing at infinity in the near-horizon basis in the large $\epsilon$ regime. Indeed, at large mode number $k$ of the Matsubara frequencies \eqref{MatsubaraSchw},
\begin{equation}
\epsilon\vert_{\omega=\omega_k}=k,
\end{equation}
and the relevant confluent Heun connection formula is the one associated to the dual regime in the gauge theory language, as studied in \cite{Bonelli:2022ten}.
We write
\begin{equation}\label{eq:connection}
\psi_{\text{up}}(z)=\mathcal{C}^{(D)}_{\text{up},\text{in}}\,\psi_{\text{in}}(z)+\mathcal{C}^{(D)}_{\text{up},\text{out}}\,\psi_{\text{out}}(z),
\end{equation}
where the expressions of the connection coefficients and the relevant details can be found in the supplemental material.

Using \eqref{eq:connection}, the near-horizon Green's function decomposes as
\begin{equation}
\widetilde{G}_R=\widetilde{G}_++\widetilde{G}_-,
\end{equation}
where
\begin{align}
\widetilde{G}_+(\omega;z,z')&=-\frac{M}{a_1}\frac{\mathcal C^{(D)}_{{\text{up}},{\text{in}}}}{\mathcal C^{(D)}_{{\text{up}},{\text{out}}}}p(z)p(z')\psi_{\text{in}}(z)\psi_{\text{in}}(z'),\label{eq:Gplus}\\
\widetilde{G}_-(\omega;z,z')&=-\frac{M}{a_1}p(z)p(z')\psi_{\text{in}}(z)\psi_{\text{out}}(z').
\label{eq:Gminus}
\end{align}
The residue sum from $\widetilde{G}_+$ closed in the UHP is a Taylor series in $X^{-1}$, while the residue sum from $\widetilde{G}_-$ closed in the LHP is a Taylor series in $X$. Altogether both contributions form a Laurent series which converge in an annular region in the complex $X$ plane. 

We start with $\widetilde{G}_+$, which contributes residues from $k\geq 0$ Matsubara poles,
\begin{equation}
\begin{aligned}
&\mathrm{res}^{(+)}_k\equiv\mathrm{res}\left[\frac{e^{-i\,\omega\,(t-t')}}{2\pi}\widetilde{G}_+(\omega, z, z'),\omega=\omega_k\right] =\\
&\mathrm{res}\left[-\frac{e^{-i\,\omega\,(t-t')}}{2\pi}\frac{M}{a_1}\frac{\mathcal{C}^{(D)}_{\text{up},\text{in}}}{\mathcal{C}^{(D)}_{\text{up},\text{out}}}p(z)p(z')\psi_{\text{in}}(z)\psi_{\text{in}}(z'),\omega=\omega_k\right].
\end{aligned}
\end{equation}
We use the asymptotic (we stress that this is a large $\omega$ approximation rather than a large $r_*$ one) expansion for the wave solutions:
\begin{equation}
\frac{e^{-i\omega\,r_*}}{\rho_{\text{in}}}\sim p(z)\psi_{\text{in}}(z),
\end{equation}
with 
\begin{equation}\label{rhofactor}
\rho_{\text{in}}=e^{- i \omega\,(2M+C) }.
\end{equation}
With the confluent Heun connection formulae provided in the supplemental material, we find the following approximation for the residues:
\begin{equation}
\begin{aligned}
\mathrm{res}^{(+)}_k
\sim\,&\frac{i(-1)^{k+1}}{2\,(k!)^2}\frac{e^{\frac{k}{4 M}\,(t-t'+r_*+r'_*-2C)}}{2\pi}\frac{k^k}{e^k}\times\\
&\frac{\Gamma\left(\frac{2 k^2+k+2 \ell (\ell+1)-2 s^2+1}{2 k}\right)}{\Gamma\left(\frac{k+2 \ell (\ell+1)-2 s^2+1}{2 k}\right)}.
\end{aligned}
\end{equation}
Using Stirling approximation, the asymptotic of the residues is given by 
\be
\mathrm{res}_k^{(+),\text{asy}} = \frac{i(-1)^{k+1}}{4\sqrt{2}\pi^2} \frac{e^{\frac{k}{4 M}\,(t-t'+r_*+r'_*-2C)} }{k}.
\ee
By summing over these, we find
\be
\sum_{k > 0} \mathrm{res}_k^{(+),\text{asy}} = \frac{i}{4\sqrt{2}\pi^2} \log\left(1+e^{\frac{1}{4 M}\,(t-t'+r_*+r'_*-2C)}\right).
\ee
This has a logarithmic branch point singularity when $e^{\frac{1}{4 M}\,(t-t'+r_*+r'_*-2C)} = -1$. Thus we find that the singularity that sets the inner radius of the $X$-plane annulus is a logarithmic singularity at \eqref{pastbounce}. This is illustrated in FIG. \ref{fig:Xplanes}.

Next we consider $\widetilde{G}_-(\omega;z,z')$, which contributes residues from $k \leq 0$ Matsubara poles,
\begin{equation}
\begin{aligned}
&\mathrm{res}^{(-)}_k\equiv\mathrm{res}\left[\frac{e^{-i\,\omega\,(t-t')}}{2\pi}\widetilde{G}_-(\omega, z, z'),\omega=\omega_k\right] =\\
&\mathrm{res}\left[-\frac{e^{-i\,\omega\,(t-t')}}{2\pi}\frac{M}{a_1}p(z)p(z')\psi_{\text{in}}(z)\psi_{\text{out}}(z'),\omega=\omega_k\right].
\end{aligned}
\end{equation}
Using the asymptotic expansions
\begin{equation}
\begin{aligned}
\frac{e^{-i\omega\,r_*}}{\rho_{\text{in}}}&\sim p(z)\psi_{\text{in}}(z),\\
\frac{e^{i\omega\,r_*}}{\rho_{\text{out}}}&\sim p(z)\psi_{\text{out}}(z),
\end{aligned}
\end{equation}
with $\rho_{\text{in}}$ as in \eqref{rhofactor} and $\rho_{\text{out}}=e^{i \omega  (2 M+C)}$, we find
\begin{equation}
\mathrm{res}_k^{(-),\text{asy}}=\frac{M}{\pi  k} e^{\frac{k (t-t'+r_*-r'_*)}{4 M}}.
\end{equation}
Summing over these, we find
\begin{equation}
\sum_{k< 0}\mathrm{res}_k^{(-),\text{asy}}=\frac{M}{\pi } \log \left(1-e^{\frac{t-t'-r_*+r'_*}{4 M}}\right)
\end{equation}
and hence a log branch point singularity at $e^{-\frac{1}{4M}\left(t-t'+r_*-r_*'\right)} = 1$, corresponding to the ingoing lightcone singularity setting the outer radius of the $X$-plane annulus. This is illustrated in FIG. \ref{fig:Xplanes}.

In this section we have thus  established the annular region of convergence for the Matsubara mode sum, between the past bouncing singularity and the ingoing lightcone singularity in the $X$ plane. In the real time domain, this corresponds to convergence between the null rays illustrated in FIG. \ref{fig:decomposition}.

\section{Discussion}
In this work we established that the null rays determining the convergent spectral decomposition of perturbations to the Schwarzschild spacetime from \cite{Arnaudo:2025uos} result from singularities at complex $t$, \eqref{futurebounce} and \eqref{pastbounce}. These singularities correspond to the propagation of the perturbation along a null geodesic which bounces from the black hole singularity.

These `bouncing singularities' are not singularities of the physical spacetime but are nevertheless responsible for setting the radius of convergence of the late time QNM expansion which, based on asymptotic analysis, is governed by the closest singularity in the complex $X=e^{-\frac{2\pi}{\beta}t}$ plane to $X=0$, as in FIG. \ref{fig:Xplanes}. The closest are the bouncing singularities which sit at $X<0$ (complex $t$), whereas the physical region is $X>0$ (real $t$). The boundary of the disk of convergence when $X>0$ is the null ray used in the decomposition \cite{Arnaudo:2025uos}. 

Our work establishes the geometrical origin for the radius  $r^\text{bounce}$ given in \eqref{rbounce}. While its value is seemingly insignificant from the perspective of the exterior geometry, we have shown its geometrical significance arises from considering propagation of null singularities in the maximally extended Schwarzschild spacetime, which set the region of convergence of the QNM expansion.

We note that bouncing singularities are also responsible for the radius of convergence of QNMs in P\"oschl-Teller. This can be seen in section IIB of \cite{Arnaudo:2025uos}, where a closed-form expression for the large $r_*, r_*'$ Green's function is given, displaying a logarithmic branch point singularity precisely at \eqref{futurebounce} with $\beta = 2\pi$.

These complex-time singularities were first identified by considering bouncing geodesics in the context of AdS/CFT, where they provide useful insight into the analytic structure of the boundary correlators. This was motivated by the possibility of probing the physical region behind the horizon using analytic structure of the correlator. It is therefore tempting to make the same connection between behind-the-horizon physics of astrophysical black holes and QNM convergence regions.

We remark also that, based on results in \cite{Pereniguez:2026avs}, building up on \cite{Mukkamala:2024dxf}, our analysis appears to cover not only vector-type gravitational perturbations, but scalar-type ones as well.

It would be interesting to investigate QNM convergence regions and the analogue of $r^\text{bounce}$ in other spacetimes such as the larger Kerr-Newman family. The geometric origin we have established in the form of bouncing singularities depend on the structure of the maximally extended spacetime. For instance, Reissner-Nordstr\"om has timelike singularities and this may lead to different convergence criteria.

\paragraph{Acknowledgements.}
\begin{acknowledgments}
We would like to thank Adrien Kuntz for discussions. We also thank Cristoforo Iossa, Robin Karlsson and IGAP Trieste for hospitality during the workshop `Black Hole Perturbations and Holography' where this work was initialised, as well as Emanuele Berti and the Simons Collaboration on Black Holes and Strong Gravity for hospitality while this work was being finalised.  PA is supported by the Royal Society grant URF\textbackslash R\textbackslash 231002, `Dynamics of holographic field theories'.
BW is supported by a Royal Society University Research Fellowship and in part by the Science and Technology Facilities Council (Consolidated Grant `New Frontiers in Particle Physics, Cosmology and Gravity'). 
\end{acknowledgments}

\bibliography{refs} 

\onecolumngrid

\section*{SUPPLEMENTAL MATERIAL}

\subsection{Confluent Heun equation}\label{app:CHeun}

The dictionary relating \eqref{RWCHeun} to the confluent Heun equation 
\begin{equation}
\begin{aligned}
\psi''(z)+\biggl(\frac{a_0^2+a_1^2+u-\frac{1}{2}}{z (z-1)}+\frac{\frac{1}{4}-a_0^2}{z^2}+\frac{\frac{1}{4}-a_1^2}{(z-1)^2}-\frac{\epsilon^2}{4}+\frac{\epsilon \mu}{z}\biggr)\psi(z)=0,
\end{aligned}
\end{equation} is\footnote{We remark that the dictionary is not unique, since for example only the squares of $a_0$ and $a_1$ are fixed. We take a concrete choice and work consistently with it.}
\begin{equation}
\begin{aligned}
a_0&=s,\\
a_1&=2i M\omega,\\
\epsilon&=-4i M\omega,\\
\mu&=2i M\omega,\\
u&=12 M^2 \omega ^2-\ell (\ell+1).
\end{aligned}
\end{equation}

We can represent the local solutions of the confluent Heun equation around the black hole horizon $z=1$ as
\begin{equation}
\begin{aligned}
\psi_{\text{in}}(z)&=z^{\frac{1}{2}-a_0}(z-1)^{\frac{1}{2}-a_1}e^{\frac{\epsilon z}{2}}e^{-\frac{\epsilon}{2}}\mathrm{HeunC}(q-\alpha,-\alpha,\delta,\gamma,-\epsilon;z-1),\\
\psi_{\text{out}}(z)&=z^{\frac{1}{2}-a_0}(z-1)^{\frac{1}{2}+a_1}e^{\frac{\epsilon z}{2}}e^{-\frac{\epsilon}{2}}\mathrm{HeunC}\bigl(q-\alpha-(1-\delta)(\epsilon+\gamma),-\alpha-(1-\delta)\epsilon, 2-\delta,\gamma,-\epsilon;z-1 \bigr),
\end{aligned}
\end{equation}
where the confluent Heun function admits a convergent expansion around $z=0$ given by
\begin{equation}\label{HeunCtaylor}
\mathrm{HeunC}(q,\alpha,\gamma,\delta,\epsilon;z) = 1 - \frac{q}{\gamma}z + \mathcal{O}(z^2),
\end{equation}
and where the Greek parameters are given by
\begin{equation}\label{greek}
\begin{aligned}
\gamma=&1-2s,\\
\delta=&1-4i M\omega,\\
\epsilon=&-4i M\omega,\\
\alpha=&4i M\omega(s-1),\\
q=&\ell(\ell+1)-s(s-1).
\end{aligned}
\end{equation}

Around the irregular singular point $z=\infty$ representing infinity, the local solutions are expressed in terms of a different function $\mathrm{HeunC}_\infty$:
\begin{equation}
\begin{aligned}
\psi_{\text{down}}(z)&=z^{-\mu}e^{\frac{\epsilon z}{2}}\mathrm{HeunC}_\infty(q,\alpha,\gamma,\delta,\epsilon;z^{-1}),\\
\psi_{\text{up}}(z)&=z^{\mu}e^{-\frac{\epsilon z}{2}}\mathrm{HeunC}_\infty(q-\gamma\epsilon,\alpha-\epsilon(\gamma+\delta),\gamma,\delta,-\epsilon;z^{-1}),
\end{aligned}
\end{equation}
where $\mathrm{HeunC}_\infty$ admits an asymptotic expansion around $z=\infty$ given by (see \cite{motygin2018evaluation}):
\begin{equation}\label{HeunCinfty}
\begin{aligned}
&\mathrm{HeunC}_\infty(q,\alpha,\gamma,\delta,\epsilon;z^{-1}) \sim 1+ \frac{\alpha ^2+\alpha  \epsilon  (-\gamma -\delta +\epsilon +1)-(q+1) \epsilon ^2}{\epsilon ^3} z^{-1} + \mathcal{O}(z^{-2}).
\end{aligned}
\end{equation}

The connection coefficients in \eqref{eq:connection} are given by
\begin{equation}
\begin{aligned}
\mathcal{C}^{(D)}_{\text{up},\text{in}}&=e^{-\epsilon/2}e^{-\frac{1}{2}\partial_{\mu}F_D}\epsilon^{\frac{1}{2}+\mu_D-\mu}\frac{\Gamma(2a_1)}{\Gamma\left(\frac{1}{2}+\mu_D-\mu+a_1\right)}e^{\frac{1}{2}\partial_{a_1}F_D}\,\epsilon^{-a_1},\\
\mathcal{C}^{(D)}_{\text{up},\text{out}}&=e^{-\epsilon/2}e^{-\frac{1}{2}\partial_{\mu}F_D}\epsilon^{\frac{1}{2}+\mu_D-\mu}\frac{\Gamma(-2a_1)}{\Gamma\left(\frac{1}{2}+\mu_D-\mu-a_1\right)}e^{-\frac{1}{2}\partial_{a_1}F_D}\,\epsilon^{a_1},
\end{aligned}
\end{equation}
where $\mu_D$ plays the role of the parameter $a$ and $F_D$ is the dual prepotential. At zero instantons, we use the relation \cite{Lisovyy:2021bkm}
\begin{equation}\label{matonelargeL}
u=-(\mu_D-\mu)\epsilon+\frac{1}{2}-a_0^2-a_1^2+2\mu_D(\mu_D-\mu),
\end{equation}
up to terms of order $1/\epsilon$, and $F_D=\mathcal{O}(\epsilon^{-1})$.
Inverting \eqref{matonelargeL} gives 
\begin{equation}
\mu_D=\mu+\frac{\frac{1}{2}-a_0^2-a_1^2-u}{\epsilon}+\mathcal{O}(\epsilon^{-2}).
\end{equation}

With these conventions, and in the large frequency approximation, the residues of the positive Matusbara frequencies can be written as
\begin{equation}
\begin{aligned}
&\mathrm{res}\left[-\frac{e^{-i\,\omega\,(t-t')}}{2\pi}\frac{M}{a_1}\frac{\mathcal{C}^{(D)}_{\text{up},\text{in}}}{\mathcal{C}^{(D)}_{\text{up},\text{out}}}p(z)p(z')\psi_{\text{in}}(z)\psi_{\text{in}}(z'),\omega=\omega_k\right]\sim\\
&\mathrm{res}\biggl[-\frac{e^{-i\,\omega\,(t-t'+r_*+r'_*)}}{2\pi\,\rho_{\text{in}}^2}\frac{M}{a_1}e^{\partial_{a_1}F_D}\epsilon^{-2a_1}\frac{\Gamma(2a_1)\Gamma\left(\frac{1}{2}+\mu_D-\mu-a_1\right)}{\Gamma(-2a_1)\Gamma\left(\frac{1}{2}+\mu_D-\mu+a_1\right)},\omega=\omega_k\biggr],
\end{aligned}
\end{equation}
which, at leading instanton order, and using that the poles come from $2a_1=-k$, equal
\begin{equation}
\begin{aligned}
&\frac{i(-1)^k}{4M\,k!}\frac{e^{-i\,\omega\,(t-t'+r_*+r'_*)}}{2\pi}\frac{M}{a_1\,\rho_{\text{in}}^2}\epsilon^{-2a_1}\frac{\Gamma\left(\frac{1}{2}+\mu_D-\mu-a_1\right)}{\Gamma(-2a_1)\Gamma\left(\frac{1}{2}+\mu_D-\mu+a_1\right)}\bigg|_{\omega=\omega_k}=\\
&\frac{i(-1)^{k+1}}{2\,(k!)^2}\frac{e^{\frac{k}{4 M}\,(t-t'+r_*+r'_*-2C)}}{2\pi}\frac{k^k}{e^k}\frac{\Gamma\left(\frac{2 k^2+k+2 \ell (\ell+1)-2 s^2+1}{2 k}\right)}{\Gamma\left(\frac{k+2 \ell (\ell+1)-2 s^2+1}{2 k}\right)}.
\end{aligned}
\end{equation}

\end{document}